\documentclass [prb,aps,superscriptaddress,showpacs,showkeys,twocolumn,letterpaper] {revtex4}

\usepackage [breaklinks] {hyperref}

\pdfoutput=1
\usepackage {graphicx}
\DeclareGraphicsExtensions {.png}
\newcommand {\nc} {\newcommand}

\nc {\xopt} {$x = 0.16$}
\nc {\xod} {$x = 0.18$}
\nc {\sixteen} {$^{16}$O}
\nc {\eighteen} {$^{18}$O}
\nc {\sixteenr} {$^{16R}$O}
\nc {\bisco} {Bi$_2$Sr$_2$CaCu$_2$O$_{8+\delta}$}

\nc {\aaf} {aaf}
\nc {\ef} {E_F}
\nc {\ek} {\epsilon ({\bf k})}
\nc {\ep} {E_p}
\nc {\gkw} {G \kw}
\nc {\ikw} {I \kw}
\nc {\ims} {\mathrm {Im}\Sigma}
\nc {\kf} {{\bf k}_F}
\nc {\kw} {({\bf k}, \omega)}
\nc {\mkw} {M \kw}
\nc {\op} {\omega_p}
\nc {\skw} {\Sigma \kw}
\nc {\res} {\mathrm {Re}\Sigma}
\nc {\tc} {T_c}
\nc {\tstar} {T^*}
\begin{document}

%@<< title page >>
%@+node:<< title page >>
\title {Unusual oxygen isotope effects in cuprates -- importance of doping}

\author {G.-H. Gweon}
\affiliation {Department of Physics, University of California, Berkeley, CA 94720} 
\affiliation {Department of Physics, University of California, Santa Cruz, CA 95064}

\author {T. Sasagawa}
\affiliation {Department of Advanced Materials Science, University of Tokyo, Kashiwa, Chiba 277-8561, Japan}
\affiliation {CREST, Japan Science and Technology Agency, Saitama 332-0012, Japan}
\affiliation {Materials and Structures Laboratory, Tokyo Institute of Technology, Kanagawa 226-8503, Japan}
 
\author {H. Takagi}
\affiliation {Department of Advanced Materials Science, University of Tokyo, Kashiwa, Chiba 277-8561, Japan}
\affiliation {CREST, Japan Science and Technology Agency, Saitama 332-0012, Japan}
\affiliation {RIKEN (The Institute of Physical and Chemical Research), Wako 351-0198, Japan}

\author {D.-H. Lee}

\author {A. Lanzara}
\affiliation {Department of Physics, University of California, Berkeley, CA 94720}
\affiliation {Materials Sciences Division, Lawrence Berkeley National Laboratory, Berkeley, CA 94720}

\date {\today}

\begin {abstract}
A recent angle resolved photoelectron spectroscopy (ARPES) study by Douglas et al.~\cite {dessau-comment} on oxygen isotope exchanged \bisco~superconductors reported an absence of isotope effect at optimal doping, questioning the previous work by us~\cite {gweon-nature}.
Here, we report a new result that sheds light on this puzzling discrepancy as well as the nature of the electron lattice interaction in the cuprates: the anomalous isotope effect at optimal doping \cite {gweon-nature}, re-confirmed here, vanishes on a mere 2 \% overdoping of holes.
This result implies a rapid change of the nature of the electron-lattice interaction near optimal doping.
We also find that the data by Douglas et al.~\cite {dessau-comment} are actually characteristic of significantly over-doped samples, not of optimally doped samples as they claimed.

\end {abstract}


\maketitle
%@nonl
%@-node:<< title page >>
%@nl

%@<< intro >>
%@+node:<< intro >>
Whether there is a critical doping concentration at which the electronic structure of hole-doped high temperature superconductor material undergoes dramatic change has been a subject of debate over the past.  For example, pump and probe experiments \cite {orenstein-prl} revealed a dramatic change of the quasiparticle recombination dynamics near optimal doping.  Other thermodynamic and transport data have also pointed toward the possibility of a critical doping concentration somewhere in the vicinity of optimal doping \cite {ando-prl, boebinger-prl}.  In this paper we shed some light on this question by examining the effect of oxygen isotope substitution (\sixteen$\rightarrow$\eighteen) on the photoemission data.  Our main conclusion is that the isotope effect is very sensitive to even a minute amount of doping in excess of the optimal doping.  This result suggests that the electron-lattice coupling undergoes a sharp change across the optimal doping.

In this paper, we extend our previous study of the oxygen isotope effect of the angle resolved photoelectron spectroscopy (ARPES) on \bisco~to slightly over-doped samples.  The new data on optimally doped samples show anomalous isotope effect in off-nodal regions, in good agreement with previous results \cite {gweon-nature, gweon-prl}.  In addition, as $x$, the amount of doped holes per unit cell of the 2-dimensional CuO$_2$ layer, increases by a mere 2 \%, the isotope effect disappears within the error bar of the experiment ($\approx \pm 10$ meV).  These results are consistent with the abrupt changes observed by other probes at similar doping values \cite {orenstein-prl, ando-prl, boebinger-prl}.

In light of these findings, we respond to a recent comment on our previous work by Douglas et al.~\cite {dessau-comment}, who reported the absence of an anomalous isotope effect for ``optimally doped'' samples.  Indeed, judging from the published data of Douglas et al.~\cite {dessau-comment} we estimate the doping level of their sample to be at least 2 \% in excess of optimal doping, which explains why they do not observe a significant isotope effect.
%@nonl
%@-node:<< intro >>
%@nl

%@<< experiment >>
%@+node:<< experiment >>
The new ARPES data reported here were obtained at beam line (BL) 7 of the Advanced Light Source (ALS) of the Lawrence Berkeley National Laboratory, using photons of energy 100 eV at temperature $\approx 25$ K\@.
The energy resolution was set to $\approx 60$ meV or $\approx 35$  meV while the angular resolution was set to $\approx 0.35^\circ$.
Note that although these energy resolution values are rather large compared to the best available energy resolution ($1 \sim 10$ meV), they are comparable to the resolution used in Ref.~\onlinecite {dessau-comment}, and are sufficient for discussing high energy peaks, the main focus of this paper, whose intrisinc widths are well above 100 meV\@.
Despite the relatively large energy resolution, the peak positions
described below, extracted from peak maxima, have an uncertainty of only $\pm 10$ meV\@.    
We also discuss (in Fig.'s 3 and 4a) data obtained at BL 10 of the ALS, as described before \cite {gweon-nature, gweon-prl}.  
It is important to note that both in the experiments reported here and in those reported in our previous work \cite {gweon-nature, gweon-prl} the two oxgyen-isotope samples were measured one right after the other in an otherwise identical experimental set-up.
%@nonl
%@-node:<< experiment >>
%@nl

\begin{figure}

\includegraphics*[width=3 in]{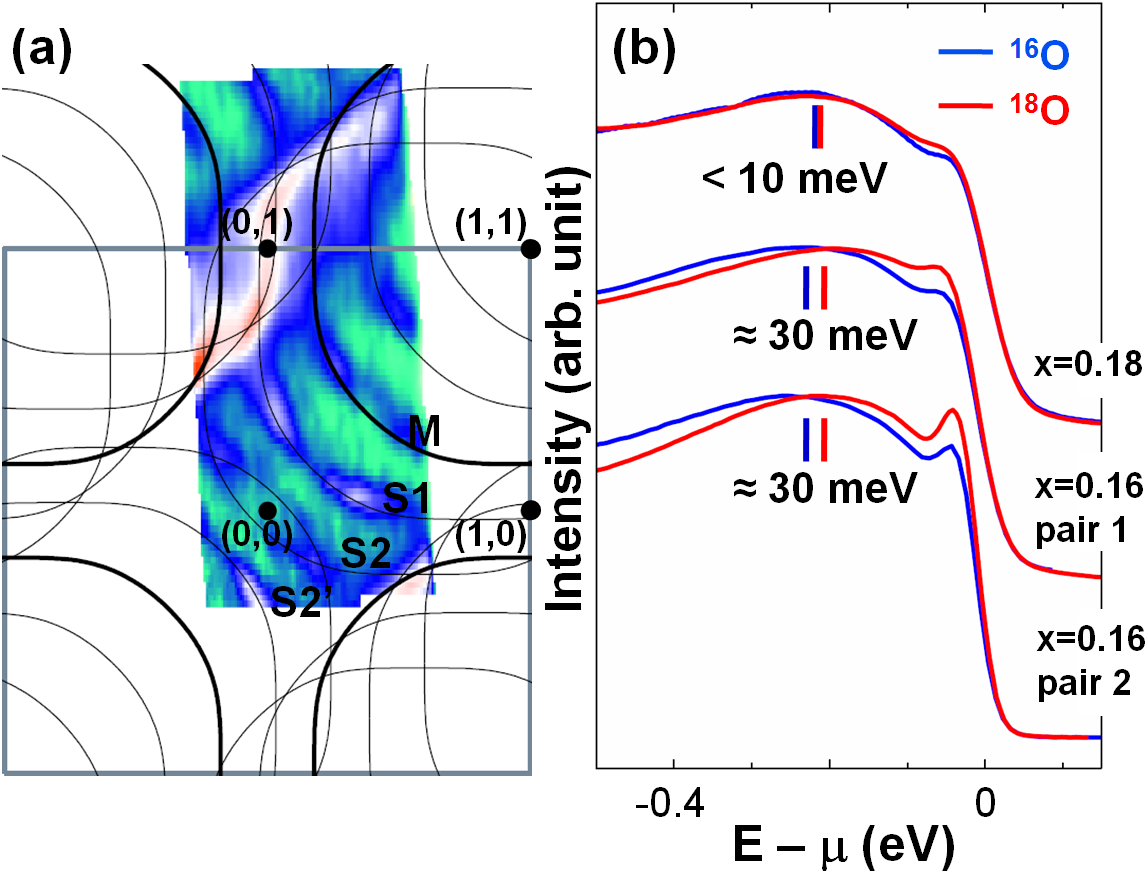}

\caption {(a) Constant energy map at the chemical potential ($\mu$) on optimally doped \bisco~sample.  The boundary for the first Brillouin zone is shown as a square, and four selected high symmetry points are explicitly indicated in unit of $(\pi/a, \pi/a)$, where $a$ is the lattice constant of the CuO$_2$ plane.  (b) Angle integrated photoemission data obtained on three different sets of samples, at optimal doping ($x=0.16$) and a slight over doping ($x=0.18$).  Data are normalized to the area under the curve for $E - \mu > -0.5$ eV\@.  For ease of view, each pair of curves has been shifted in intensity, which approaches zero at far right end of the energy axis for all curves.}

\label {fig-AIPES}

\end{figure}

%@<< angle integrated data >>
%@+node:<< angle integrated data >>
Fig.~\ref {fig-AIPES} shows data obtained on optimally doped samples and over-doped samples, with $x \approx 0.16$ and $x \approx 0.18$ respectively.  As we shall discuss later, although usually it is difficult to estimate the absolute doping level to an accuracy better than 1 \% (i.e.~0.01 for $x$), ARPES is sensitive to a doping change of 1 \%.

Fig.~\ref {fig-AIPES}(a) shows a constant energy map at the chemical potential ($\mu$) for $x = 0.16$ sample, giving an overview of the Fermi surface.  This map fits well with a tight binding parameterization \cite {markiewicz}, whose Fermi surface and its replica are plotted as solid lines.  The high photon energy used here allows to enhance the super-structure replicas ``S1'' and ``S2,'' first order and second order replicas due to the well-known surface reconstruction in the Bi-oxide layer.  
As we will discuss extensively below, these are key features for determining the doping with high accuracy ($\ll 0.01$).
It is important to note that, because of the geometry and photon energy used in this experiment, the ARPES signal is a local minimum along the nodal direction, consistent with previous findings using the same geometry \cite {lanzara-old}.  
This situation is very different from the previous one \cite {gweon-nature} in which the nodal intensity was a local maximum.  
For this reason, in this paper we focus on the isotope effect only in the off-nodal region, where we have sufficient signal to noise ratios to discuss large isotope effects  \cite {gweon-nature}.

%@<< fig-AIPES >>
%@+node:<< fig-AIPES >>
%@nonl
%@-node:<< fig-AIPES >>
%@nl

Fig.~\ref {fig-AIPES}(b) shows angle integrated data as a function of the oxygen isotope for three pairs of isotope-exchanged samples and two different doping values.  The upper two sets of data were obtained with a total energy resolution of $\approx 60$ meV while the bottom set of data with a total energy resolution of $\approx 35$ meV\@.  These curves present the main result of this paper: the isotope induced changes are strongly doping dependent.  
Note first that there is a clear isotope-induced shift of the broad high energy bump at $\approx -0.2$ eV by $\approx 30$ meV\@ at optimal doping ($x = 0.16$), in agreement with previous reports \cite{gweon-nature, gweon-prl}.
Surprisingly the isotope shift is reduced to less than 10 meV when $x$ increases by a mere $\Delta x = 0.02$. 

Douglas et al.~\cite {dessau-comment} have argued that the isotope-induced shift observed at optimal doping is an artifact caused by errors in sample alignment.
We have chosen to compare angle integrated data \cite {note-momentum-sum}, as they are clearly immune to such systematic errors.
Fig.~\ref {fig-AIPES} shows that the isotope induced shift is indeed an intrinsic effect and that 
the disparity between our previous data and those of Ref.~\onlinecite {dessau-comment} can not be attributed to sample mis-alignment.
In any case, uncertainty in sample alignment has been checked to fall within 0.5$^\circ$, and we now report angle resolved data, which reflect well the above finding, the small uncertainty in sample alignment notwithstanding.
%@nonl
%@-node:<< angle integrated data >>
%@nl

%@<< EDCs >>
%@+node:<< EDCs >>
%@<< fig-EDCs >>
%@+node:<< fig-EDCs >>
\begin{figure}

\includegraphics*[width=3.3 in]{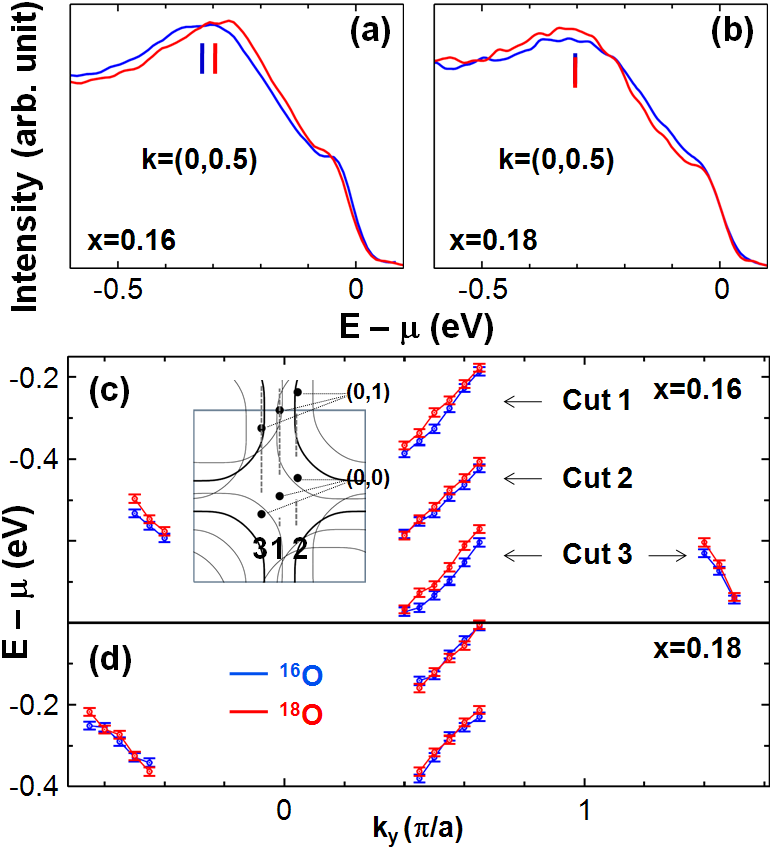}

\caption {(a, b) EDCs for momentum value (0, 0.5), showing a finite isotope shift for optimally doped sample (a) and no shift for over-doped sample (b).  (c) EDC peak positions compiled for three cuts, arising from main (cut 1) and superstructure replicas (cuts 2,3) for $x = 0.16$.  When momentum values for cuts 2,3 are normalized, by subtracting relevant superstructure wave vectors, it can be seen that all of these three cuts are defined on the same vertical line, passing through points (0, 0) and (0, 1), as shown in the inset.  The horizontal axis of this plot is defined by this normalized momentum value.  Note also that the graphs for cuts 2,3 are shifted in the increments of -0.2 eV for easy view of data, but it should be understood that each cut has an approximate energy range -0.2 eV to -0.4 eV\@.  (b) EDC peak positions for $x = 0.18$, plotted in the same manner as in (c).}

\label {fig-EDCs}

\end{figure}
%@nonl
%@-node:<< fig-EDCs >>
%@nl

Fig.'s \ref {fig-EDCs}(a,b) show representative EDCs at (0, 0.5) (throughout this paper, we use  $\pi/a$ as unit of momentum, where $a$ is the lattice constant of the CuO$_2$ plane).
Note that the $x = 0.16$ sample shows a clear isotope shift, $\approx$ 30 meV, while the $x = 0.18$ sample shows no isotope shift within the uncertainty $\pm 10$ meV\@.
Panels c,d show the effect of isotope substitution on the position of high energy peak in the momentum-energy plane.
The cuts investigated here are along the (0,0) to (0,1) direction, where we expect large isotope effect \cite{gweon-nature, gweon-prl}.  
Note that the peak positions of EDCs for optimal doping show a definite isotope-induced energy shift as all of the red symbols are above the blue symbols, with an exception of one data point in cut 2.
The shift is large, up to 40 meV, in good agreement with our previous work \cite {gweon-nature, gweon-prl}.
The effect is observed also for super-structure replicas (cuts 2,3), for positive and negative $k_y$ values (cut 2), and for {\bf k} values in the first and the second Brillouin zones (cut 3).
That is, this anomalous isotope effect observed has the correct expected symmetry.
Also, it can be noted that as the binding energy is increased beyond 300 meV, the isotope effect decreases.  Thus, the ``active energy range'' of the anomalous isotope effect is $200 - 300$ meV, consistent with our previous result \cite {gweon-nature}.  In contrast, the over-doped sample shows no isotope effect within the error bar, as shown in panel d \cite {width}.
%@nonl
%@-node:<< EDCs >>
%@nl

%@<< nodal cut >>
%@+node:<< nodal cut >>
The data presented above disagree with the data of Douglas et al.\cite {dessau-comment}, assuming their sample is optimally doped.  In view of the sensitivity of the isotope effect to a minute doping change, we decided to take a closer look at the doping value of their samples.

\begin{figure}

\includegraphics*[width=2.4 in]{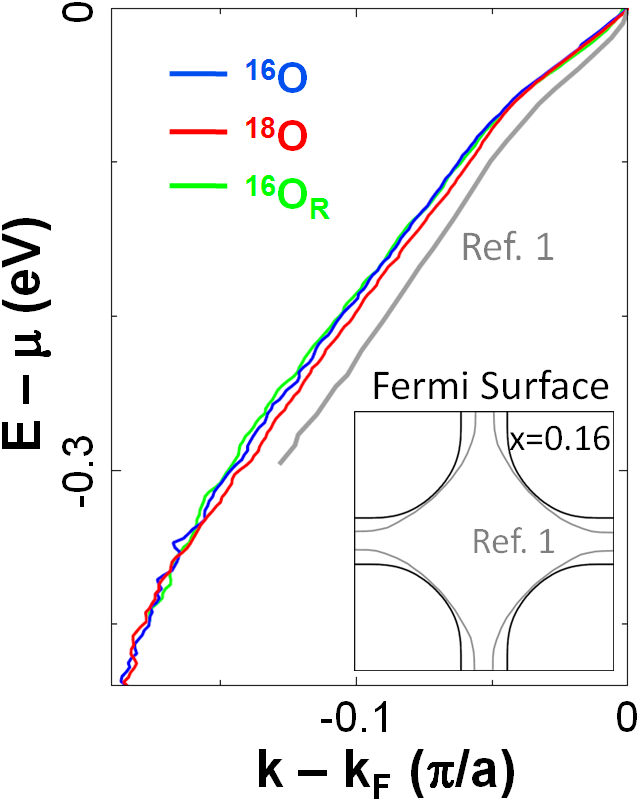}

\caption {The nodal cut data of Ref.~\onlinecite {dessau-comment} compared with our nodal cut data.  Also we show line shape fit results for a more extended energy range than we presented previously \cite {gweon-nature, gweon-prl}.  Inset compares our Fermi surface with the Fermi surface of Ref.~\onlinecite {dessau-comment}.}

\label {fig-nodal-cut}

\end{figure}

Fig.~\ref {fig-nodal-cut} shows that the doping of samples in Ref.~\onlinecite {dessau-comment} is significantly different from ours.  This can be seen from the comparison of the dispersions along the nodal cut, as shown in the main panel.
In an earlier comprehensive study \cite {zhou-nature} it has been shown that   
the ratio of the high energy ($\mu - E = 100 - 300$ meV) group velocity to the low energy ($\mu - E < 50$ meV) group velocity is a sensitive indicator of the doping level, reflecting the fact that the ARPES ``kink'' becomes milder as the doping increases. 
This ratio for the data of Ref.~\onlinecite {dessau-comment} ($\approx 1.5$) is smaller than that for our data ($\approx 1.7$) by $\approx 0.2$, which corresponds \cite {zhou-nature} to an over-doping by $\Delta x = 0.02 \sim 0.03$.

An over-doping of the samples in Ref.~\onlinecite {dessau-comment} is also apparent from both the Fermi surface and the superconducting gap size.  As the inset of Fig.~\ref {fig-nodal-cut} shows, their Fermi surface is very different from the well-known optimal doping Fermi surface and, actually, looks like that of an overdoped sample \cite {feng-prl} with $x$ at least 0.22. 
Similarly, the absence of a clear superconducting gap in their off-nodal dispersion data \cite {dessau-comment} contradicts the well-known behavior \cite {gweon-nature, gweon-prl} at optimal doping  and is also indicative of over-doping.

While the above observations make it certain that Douglas et al.'s samples are overdoped, some differences in data remain to be understood.  For instance, note that the main cause of the difference in velocity ratio between their data and our data in Fig.~\ref {fig-nodal-cut} is the rather large low energy group velocity in Douglas et al.'s data as compared to our data as well as other data in the literature \cite {zhou-nature, kaminski-prl, valla-science, johnson-prl}, while the high energy group velocity is almost the same as that of our data.  This is quite curious in view of the well-known fact \cite {zhou-nature} that the high energy group velocity has a much greater dependence on doping, and may require a re-examination of the overall momentum scale.

%@<< fig-nodal-cut >>
%@+node:<< fig-nodal-cut >>
%@nonl
%@-node:<< fig-nodal-cut >>
%@nl

Lastly, note that the fact that our observed isotope effect in Fig.~\ref {fig-nodal-cut} vanishes as energy increases beyond 300 meV also makes it difficult to ascribe such an effect to a sample misalignment or a doping difference between isotope samples.  In addition, the latter possibility is ruled out firmly by the analysis that we now present.
%@nonl
%@-node:<< nodal cut >>
%@nl

\begin{figure}

\includegraphics*[width=3.3 in]{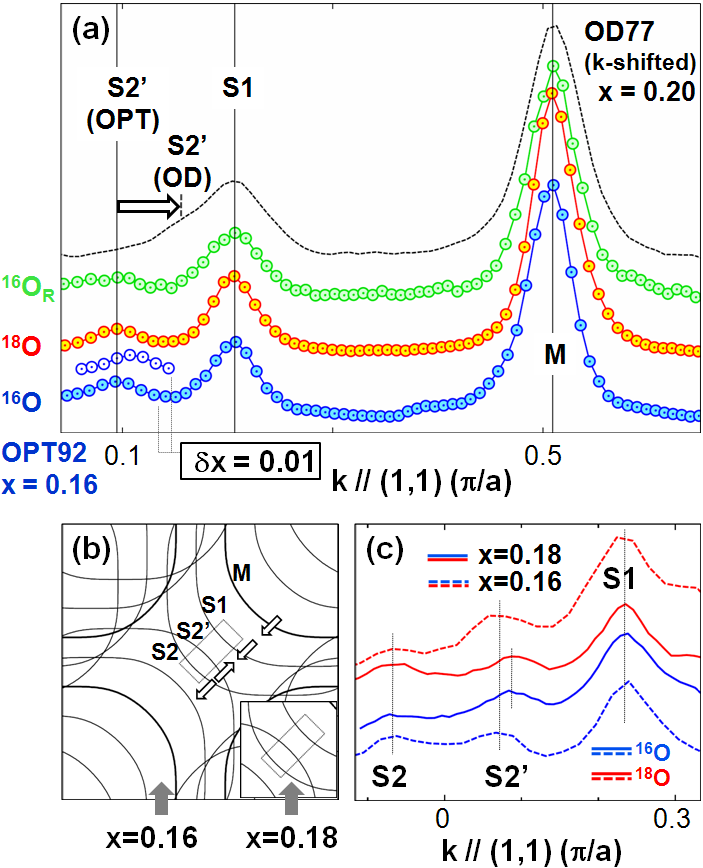}

\caption {(a) Momentum distribution curves at constant energy (35 meV below $\mu$ in order to include the S2$^\prime$ feature nicely within the cut) for optimally doped samples and a reference over-doped sample \cite {note-underdoping}.  (b)  Tight binding Fermi surface and its superstructure replicas for $x = 0.16$ and $x = 0.18$ (inset).  The same diagonal rectangle, enclosing central regions of S1, S2 and S2$^\prime$, is indicated in both plots.  Within that rectangle, note subtle but clear relative position changes of S1, S2 and S2$^\prime$ on over-doping, i.e.~S1 and S2$^\prime$ move closer while S2 and S2$^\prime$ move farther apart.  (c)  ARPES intensity at $\mu$, momentum-integrated over the short axis of the rectangle in (b), for statistics, and plotted as a function of momentum along the long axis of that rectangle.  The doping-induced changes expected from (b) are clearly observable.  For this plot, S1 and S2 features for different doping values are aligned by a small momentum shift.}

\label {fig-doping}

\end{figure}

%@<< doping >>
%@+node:<< doping >>
Fig.~\ref {fig-doping} presents the detail of how we characterize the hole doping.   It is important to note that in order to obtain the absolute $x$ value from the Fermi surface, the ARPES data in the entire Brillouin zone is required.  This is done as shown in Fig.~\ref {fig-AIPES} and our previous work (supplementary information) \cite {gweon-nature}.  However, one should keep in mind that the absolute value of $x$ is made uncertain by line shape issues, such as broad line shapes, (pseudo) gaps, bi-layer splitting, and photon energy dependence of spectral functions, all of which become increasingly important as momentum approaches the anti-nodal region.  Thus, it is not surprising that the value of $x$ typically has an error bar of magnitude $\pm 0.01$ or larger \cite {valla-condmat}.

%@<< fig-doping >>
%@+node:<< fig-doping >>
%@nonl
%@-node:<< fig-doping >>
%@nl

On the other hand, it is possible to compare the {\em relative} values of doping between two samples, by concentrating on the nodal region, where these problems are minimal.  Fig.~\ref {fig-doping}(a) compares the intensity profile along the nodal cut for our three different optimally doped samples used in Fig.~\ref {fig-nodal-cut}.  Also included is the data for a reference sample with a high over-doping \cite {note-underdoping}.  
As the doping $x$ increases the Fermi surface moves toward the $\Gamma$ point.
The key observation is that, as shown by white arrows in panel b, a doping change causes M, S1, and S2 to move in the same direction while it causes S2$^\prime$ to move in the opposite direction.
Thus, it is possible to judge the relative change of doping quite accurately by focusing on spectral peaks, e.g.~peaks from S1 and S2$^\prime$, in a narrow angular range near the normal emission -- the least complicated procedure.
For instance, peaks from S1 and S2$^\prime$ get closer, while peaks S2 and S2$^\prime$ move farther apart, as doping increases (see white arrows in a, b).
Panel a illustrates how the two peaks (at binding energy 35 meV) from S1 and S2$^\prime$ almost merge with each other for $x = 0.20$, while the position of S2$^\prime$ is still discernible as a shoulder.  Using this as reference, it is then possible to estimate how much the S2$^\prime$ peak would have moved when the doping increases by $0.01$.  This movement, in good agreement with the tight binding parametrization \cite {markiewicz}, is indicated in the lower left between the two curves for \sixteen~and \eighteen.  
It is obvious that all three optimally doped samples shown have the same doping with an error bar well within 0.01.  
Fig.'s~\ref {fig-doping}(b,c) show a similar analysis for the data of Fig.'s 1,2.  
Here, the evolution of the Fermi surface from $x = 0.16$ to $x = 0.18$ is again clearly reflected in the relative position changes of the S1, S2, and S2$^\prime$ features.
%@nonl
%@-node:<< doping >>
%@nl

%@<< what it all means >>
%@+node:<< what it all means >>
That a mere 2 \% change of doping level can cause a significant change in the response to isotope exchange may be at first quite surprising.  On the other hand, 
such spectroscopic sensitivity to doping has a precedent in another strongly correlated material, namely the manganites.  There, a recent study \cite {zou} reports that a doping change by 0.02 causes a qualitative change in the electronic structure, a complete disappearance of the quasi-particle peak, again indicating that a seemingly insignificant doping change can be very crucial.

In conclusion, we presented data on optimally doped and slightly overdoped samples of oxygen isotope exchanged \bisco~samples.  Surprisingly the anomalous isotope effect observed in optimally doped samples disappear in overdoped samples.  This study prompts a more careful study of the electronic structure near the optimal doping, where the mystery of the high temperature superconductivity seems to hide in more veils to be removed.
%@nonl
%@-node:<< closing >>
%@nl

%@<< acknowledgments >>
%@+node:<< acknowledgments >>
\begin {acknowledgments}

This work was supported by the Director, Office of Science, Office of Basic Energy Sciences, of the U.S. Department of Energy under Contract No. DE-AC03-76SF00098 and by the National Science Foundation through Grant No. DMR-0349361.

\end {acknowledgments}
%@nonl
%@-node:<< acknowledgments >>
%@nl

%@<< biblio >>
%@+node:<< biblio >>

%@nonl
%@-node:<< biblio >>
%@nl


\begin{thebibliography}{99}
\bibitem {note-momentum-sum} The momentum summation was generally not over the entire Brillouin zone, but it was effectively so considering the superstructure replicas and the approximate tetragonal symmetry.
\bibitem {width} In this paper, isotope effects on peak positions are our major concern.  However, note that there are isotope induced peak sharpenings observed in the data of Figures 1,2, consistent with our previous study \cite {gweon-prl}.
\bibitem {note-underdoping}  A reference sample with a high under-doping ($x \approx 0.1$) was also examined for this purpose with the same result.
\end{thebibliography}
\end{document}